\documentclass[12pt,preprint]{aastex}

\newcommand{\be}{\begin{equation}}
\newcommand{\ee}{\end{equation}}

\newcommand{\msun}{{M}_{\sun}}

\slugcomment{Printed at \today}
\slugcomment{Accepted by ApJ}
\shorttitle{Rapidly Rotating Black Holes in FR Is} \shortauthors{Wu, Q., et al. }

\begin{document}

\title{Evidence for Rapidly Rotating Black Holes in FR I Radio Galaxies}

\author{Qingwen Wu\altaffilmark{1}, Xinwu Cao\altaffilmark{2}, and Ding-Xiong Wang\altaffilmark{1}}

\altaffiltext{1}{Department of Physics, Huazhong University of Science and Technology,
 Wuhan 430074, China; Email: qwwu@mail.hust.edu.cn; dxwang@mail.hust.edu.cn}

\altaffiltext{2}{Key Laboratory for Research in Galaxies and Cosmology,
 Shanghai Astronomical Observatory, Chinese Academy
of Sciences, Shanghai, 200030  China; Email: cxw@shao.ac.cn}

\begin{abstract}
We investigate the correlation between 151 MHz radio luminosity,
$L_{\rm 151\ MHz}$, and jet power, $P_{\rm jet}$, for a sample of
low-power radio galaxies, of which the jet power is estimated from
X-ray cavities. The jet power for a sample of FR I radio galaxies is
estimated with the derived empirical correlation. We find that
$P_{\rm jet}/L_{\rm Edd}$ is positively correlated with $L_{\rm
X}^{2-10\rm\ {keV}}/\it L_{\rm Edd}$ for FR Is, where $L_{\rm Edd}$
is the Eddington luminosity and $L_{\rm X}^{2-10\rm\ {keV}}$ is 2-10
keV X-ray luminosity. We calculate the jet power of a hybrid model,
as a variant of Blandford-Znajek model proposed by Meier, based on
the global solution of the advection-dominated accretion flow (ADAF)
surrounding a Kerr black hole (BH). Our model calculations suggest
that the maximal jet power is a function of mass accretion rate and
the black hole spin parameter $j$. The hard X-ray emission is
believed to be mainly from the ADAFs in FR Is, and the mass
accretion rate is therefore constrained with the X-ray emission in
our ADAF model calculations. We find that the dimensionless angular
momentum of BH $j\gtrsim0.9$ is required in order to reproduce the
observed relation of $P_{\rm jet}/L_{\rm Edd}-L_{\rm X}^{2-10\rm\
{keV}}/\it L_{\rm Edd}$ for FR Is. Our conclusion will be
strengthened if part of the X-ray emission is contributed by the
jets. Our results suggest that BHs in FR I radio galaxies are
rapidly spinning, which are almost not affected by the uncertainty
of the black hole mass estimates.

\end{abstract}


\section{Introduction}

The currently most favored jet formation mechanisms include the
Blandford-Znajek (BZ) process \citep{bz77} and the Blandford-Payne
process \citep{bp82}. In the BZ process, energy and angular momentum
are extracted from a rotating black hole (BH) and transferred to a
remote astrophysical load by open magnetic field lines. In the BP
process, the magnetic fields threading the disk extract energy from
the rotation of the accretion disk itself to power the jet/outflow.
The so-called hybrid model, as a variant of BZ model, was proposed
by \citet{me99}, which combined the BZ and BP effects through the
large-scale magnetic fields threading the accretion disk outside the
ergosphere and the rotating plasma within the ergosphere.
The recent MHD simulations showed that both BH
spin and accretion process may play important roles in jet formation
\citep*[e.g.,][]{mg04,hi04,de05,hk06}, which seem to support the
hybrid model.

Recent high resolution $Chandra$ observations of the galaxy clusters and giant
elliptical galaxies have revealed prominent X-ray surface brightness depressions
corresponding to cavities or bubbles created by AGN activities
\citep*[e.g.,][]{fa00,bi04,al06,ra06}. The X-ray cavities were found to be cospatial
with the radio lobes in nine nearby low-power radio galaxies, which suggest that
these cavities are most likely inflated by the interaction of radio jet and
surrounding hot gas \citep*[e.g.,][]{mc00,al06}. Therefore, the X-ray cavities
provide a direct measurement of the mechanical energy released by the jet through the
work done on the hot gas surrounding them. Measurements of this energy, combined
with measurements of the time-scale required to inflate the cavities, can be used
to estimate the jet power \citep*[e.g.,][]{al06}.

Low-power Fanaroff-Riley I radio galaxies (FR Is) are believed to be
the BL Lac objects with relativistic jet misaligned to our line of
sight, and high-power FR IIs correspond to misaligned radio quasars
\citep*[][]{up95}. The Eddington ratios $L_{\rm bol}/L_{\rm Edd}$ of
FR I/BL Lacs are systematically lower than those of FR II/radio
quasars, with a rough division at $L_{\rm bol}/L_{\rm Edd}\sim0.01$,
which implies the accretion mode in FR I/BL Lacs may be different
from that of FR II/radio quasars \citep*[e.g.,][]{gc01,xu09}. Low
mass accretion rates $\dot{m}$
 may lead to the accretion flows to be advection
dominated \citep*[e.g.,][]{ny94}, and such hot, optically thin,
geometrically thick advection-dominated accretion flows (ADAFs) are
suggested to be present in low-luminosity AGNs \citep*[LLAGNs, e.g.,
FR Is, low luminosity Seyferts etc.,][ and references
therein]{wu07,yu07,nm08}.

The BH spin remains one of the most intriguing aspects of
astrophysics. At present, only a few observations allow spins to be
estimated for supermassive BHs, which are based on the fitting of
the `reflection' component of iron line around 6.4 keV
\citep*[e.g.,][]{wi01,fa02}. The spin energy of the BH is thought to
play an important role in powering the large-scale jets from AGNs
based on the BZ process or the hybrid model of BZ and BP processes.
In last few years, many groups tried to investigate the BH spins of
radio galaxies from their jet power based on different jet formation
models.  \citet*[][]{ne07} proposed that the BHs in the elliptical
galaxies should be rapidly rotating in order to drive powerful jets
to heat the intracluster medium and quench cooling flows. They
constrained the BH spins for nine nearby elliptical galaxies by
comparing the jet power estimates with the BZ model and the hybrid
model, where the self-similar structure of ADAF and Bondi accretion
rates are used \citep*[see][for more details]{ne07}.
\citet*[][]{mc10} further investigated the roles of the BH spin and
accretion in generating powerful AGN outbursts in the cores of 31
clusters or galaxies based on the model of \citet*[][]{ne07}.
However, they found that Bondi accretion from hot atmospheres is
generally unable to fuel these powerful AGNs and other fuel supply
should be important \citep*[see][for a similar conclusion]{wu07}.
Therefore, it is difficult to place strong constraints on the spin
parameter based on the jet power and Bondi accretion rates of these
radio galaxies.  \citet*[][]{da09} showed that the BH spins
 of low power radio sources may range from about 0.1 to 0.8, where their results are based on
 the two unclear assumptions: (1) the magnetic field strength is equal to approximately
 the Eddington magnetic field strength; (2) the field strength is proportional to the
 BH spin. \citet*[][]{wu08} proposed that the dividing line of FR I/II dichotomy can be well reproduced
 if the BH is rapidly rotating and the putative dimensionless accretion rate
 $\dot{m}\simeq0.01$ is adopted, where the accretion rate $\dot{m}\simeq0.01$ is roughly
 consistent with that of the critical rate for the accretion mode transition from a standard
 disk to an ADAF.

One of the difficulties to constrain the BH spin is that the mass
accretion rate $\dot{m}$ and the black hole spin parameter $j$ are
degenerate in jet formation models, which means that the black hole
spin parameter cannot be well constrained even after the jet power
is measured. The purpose of this paper is to constrain the BH spin
parameters for a sample of FR I radio galaxies, in which the mass
accretion rates are constrained with their nulear X-ray emission.
Most previous work on the jet power extracted from ADAFs were based on the self-similar structure of
 ADAFs \citep*[e.g.,][]{ne07,mc10}. In this work, we will calculate the jet power based on the global solution of ADAF
 surrounding Kerr BHs, since that the self-similar solution can well reproduce the global solution
 at large radii, while it deviates significantly near the BHs.  A $\rm \Lambda CDM$ cosmology with
$H_{0}=70\ \rm km\ s^{-1}\ Mpc^{-1}$, $\Omega_{\rm M}=0.27$, and
$\Omega_{\Lambda}=0.73$ is adopted in this work.

\section{Sample}
Our sample consists of two parts. The first part is the low-power
radio sources with jet power estimated from their X-ray cavities
(part I). In the literature, X-ray cavities have been divided
into two categories: radio-filled cavities and radio-ghost cavities,
 depending on the presence or absence of bright 1400 MHz radio
emission in the cavities \citep*[e.g.,][and references
therein]{bi08}. In this work, we only consider the cavities with
relatively strong radio emission(radio-filled cavities, e.g., FR I
and
 low-power FR II), where the particle injection is still occurring,
 and ghost cavities have been excluded due to the conjecture that they are
 only the relics of earlier outbursts whose radio emission has
 faded and it is no longer an indicator of current jet power \citep*[e.g.,][]{bi08}.
Twelve sources are selected from \citet*[][]{mh07} and
\citet*[][]{ca10} (see Table 1). Ten of them are FR I or FR I-like
sources, and two low-power FR IIs (3C 388 and 3C 405) are also
included. These two low-power FR IIs stay around the dividing line
of FR dichotomy in $M_{\rm BH}-P_{\rm jet}$ plane \citep*[][]{wu08},
and, therefore, they are smoothly connected to high-power FR Is. The
empirical correlation between their radio emission and jet power can
be established with this sample. We note that cavity buoyancy ages,
$t_{\rm buoy}$, are used to estimate the jet power in
\citet*[][]{ca10}, while the sound speed ages, $t_{\rm cs}$, are
used in \citet*[][and references therein]{mh07}. To be consistent
with each other, we use the buoyancy age in this work, and $t_{\rm
buoy}$ was estimated from $t_{\rm cs}$ with the mean value of the
ratio $t_{\rm cs}/t_{\rm buoy}=0.65$ \citep*[][]{bi04}. The second
part of FR I sample is primarily selected from the 3CR sample of
radio galaxies with $z<0.3$ \citep*[][]{bu10}, and some other
low-redshift FR Is in the literature are also included (part II). We
only select the FR Is with observed nuclear X-ray luminosities and
estimated BH masses (part of them the jet power is estimated by
using the empirical relation in this work). The high
resolution data of $Chandra$ is preferentially adopted to estimate
the nuclear X-ray emission, and several sources with X-ray emission
observed from $XMM$-$Newton$, $ASCA$, and $BeppoSAX$ were also
included (see Table 1). Because the X-ray data were acquired with a
variety of different instruments and analyzed with different
techniques, we convert all the luminosities to one standard
bandpass, 2-10 keV, using the available power-law flux at given
waveband and the best-fit spectral slope. Some sources with only
detected upper limits for the nuclear
 X-ray emission have not been included. Our final sample
includes 33 FR I radio galaxies (see Table 1).

\section{ADAF and Jet model}

We use the ADAF-jet model surrounding a Kerr BH to investigate the
BH spins of FR Is with their accretion/jet power. Only the main
features of the model are described here, and the details of which
can be found in \citet*[][and references therein]{wu08}. We employ
the approach suggested by \citet*[][]{man00} to calculate the global
structure of the ADAF in the general relativistic frame, which
allows us to calculate the structure of an ADAF surrounding either a
spinning or a nonspinning BH. All radiation processes (Synchrotron,
Bremsstrahlung and Compton scattering) are included consistently in
the calculations of the ADAF structure. The global structure of an
ADAF surrounding a BH spinning at a rate $j$ with mass $M_{\rm BH}$
can be calculated with proper outer boundaries, if the parameters
$\dot{m}$, $\alpha$, $\beta$ and $\delta$ are specified
\citep*[e.g.,][]{man00}. The parameter $j=J/GM_{\rm BH}^{2}c^{-1}$
is the dimensionless angular momentum of the BH ($J$ is the angular
momentum of the BH), and $\dot{m}=\dot{M}/\dot{M}_{\rm Edd}$ is the
dimensionless accretion rate (the Eddington accretion rate is
defined as $\dot{M}_{\rm Edd}=1.4\times10^{18}M_{\rm BH}/\msun \rm\
g \  s^{-1}$).  The viscosity parameter $\alpha$ in ADAF models is
supposed to be within a very narrow range
 $\alpha\simeq0.1-0.3$, which is supported either by the MHD
 simulations or the observations \citep*[e.g.,][and the references therein]{nm08}.
The magnetic parameter $\beta=P_{\rm g}/P_{\rm m}$ in our
calculations is not an independent parameter but relate to $\alpha$
as $\beta\simeq(0.55-\alpha)/\alpha$, as suggested by the MHD
simulations \citep*[e.g.,][]{ha95}. The parameter $\beta\simeq1-5$
for the typical value of $\alpha\simeq0.1-0.3$. The most poorly
constrained parameter is $\delta$, describing the fraction of the
turbulent dissipation that directly heats the electrons in the flow.
\citet*[][]{sh07} found that the parameter $\delta$ may be in range
of $\sim0.01-0.3$ based on the simulations, depending on the model
details. Recent ADAF models typically require $\delta\sim0.3$ in
order to fitting the spectra of Sgr A$^{*}$ and other LLAGNs
\citep*[e.g.,][and the references therein]{yu06}. The
field-enhancing effect caused by frame dragging is also considered
\citep[][]{me01}. The amplified magnetic field related to the
magnetic field produced by the dynamo process in the ADAF can be
expressed as $B=gB_{\rm dynamo}$, where $g=\Omega/\Omega^{'}$ is the
field enhancing factor, the disk angular velocity $\Omega$ is a sum
of its angular velocity relative to the local metric $\Omega^{'}$
plus the angular velocity of the metric itself in the
Boyer-Lindquist frame $\omega\equiv-g_{\phi t}/g_{\phi\phi }$, i.e.,
$\Omega=\Omega^{'}+\omega$. We can calculate the spectrum of the
accretion flow based on the global structure of ADAF. In the spectral
calculations, the gravitational redshift effect is considered, while
the relativistic optics near the BH is neglected \citep*[e.g.,][]{man00}.

 To evaluate the jet power extracted from the inner region of ADAF, we adopt the hybrid model
 proposed by \citet*[][]{me01}, where both the BZ and BP mechanisms were incorporated. The total
 jet power is given by
 \be P_{\rm jet} = B_{\rm p}^{2}R^{4}\Omega^{2}/32c,
\ee where $R$ is the characteristic size of jet formation region,
the poloidal magnetic field $B_{\rm p}\simeq gB_{\rm dynamo}$
\citep*[see][for the details]{me01}. Following the work by
\citet*[][]{ne07}, all the quantities are evaluated at the innermost
marginally stable orbit of the disk $R=R_{\rm ms}$, which are
roughly consistent with the jet launching regions in the numerical
simulations \citep*[e.g,][]{hi04}.

\section{Results}
\subsection{Jet Efficiency of ADAF Surrounding a Spinning BH}

We calculate the jet power extracted from the inner region of the
ADAF surrounding a rotating BH. We define the jet efficiency as
$\eta_{\rm jet}=P_{\rm jet}/\dot{M}c^{2}$, and the relations between
the jet efficiency, $\eta_{\rm jet}$, and the BH spin parameter $j$
calculated with different values of the ADAFs are shown in Figure 1.
We find that the jet efficiency is insensitive to the values of the
model parameters: $\alpha$ and $\delta$. The jet efficiency in our
numerical models is roughly consistent with those derived in the MHD
simulations (see the filled circles and squares in Figure 1)
\citep*[][]{de05,hk06}. The viscosity parameter $\alpha=0.3$ is
adopted in the following calculations due to the value of the
magnetic parameter $\beta$ in this case being roughly consistent
with that of MHD simulations \citep*[][]{hi04}.

\subsection{Estimation of the Jet Power}

The relation between the jet power, $P_{\rm jet}$, and 151 MHz radio
luminosity, $L_{\rm 151\ MHz}$, for the low-power radio galaxies
with estimated jet power from their X-ray cavities (part I in
sample) is plotted in Figure 2. A very tight positive correlation
between $P_{\rm jet}$ and $L_{\rm 151\ MHz}$ is found, and the
best-fitted linear relation is
  \be
  \log P_{\rm jet}=0.52(\pm0.04)L_{\rm 151\ MHz} + 22.31(\pm1.75),
 \ee
with an intrinsic scatter of $\sigma=0.23$ dex, which is
derived from the X-ray cavities with quite strong radio
 emission (most of them are FR Is) and radio-ghost cavities have not
been included. Therefore, this relation should be more accurate in
 estimates the jet power for the sources with relatively strong radio
 emission (e.g., FR Is in this work). We further estimate
the jet power for other 23 FR Is (part II in sample) from their 151
MHz radio luminosities with this empirical correlation (Eq. 2) (the
data are listed in Table 1).

\subsection{Estimation of the BH Mass}

The relation between the host galaxy absolute magnitude and
BH mass \citep*[e.g.,][]{mc02} was usually used to estimated the BH
mass of nearby low power radio sources. At present, the BH mass
derived from the velocity dispersion of the host bulge,
$\sigma_{*}$, is believed to be more accurate based on the empirical
correlation of $M_{\rm BH}-\sigma_{*}$ relation
\citep*[e.g.,][]{ge00,tr02},
 \be M_{\rm BH}=10^{8.13}\left(\frac{\sigma_{*}}{200\ \rm km\ s^{-1}}
\right)^{4.02} \msun, \ee with an intrinsic scatter $\simeq0.3$ dex. We estimate the
BH mass of FR Is from their velocity dispersions of host bulges
\footnote{The velocity dispersions
 were obtained from the HyperLEDA online database: http://leda.univ-lyon1.fr.}
 or select the mass from the literature with
 the same method. The BH masses of some FR Is estimated from other
  methods are also selected from the literature if there are no
observational data of $\sigma_{*}$ (see Table 1).

\subsection{Constraints on the BH spins for FR Is}

The relation between $P_{\rm jet}/L_{\rm Edd}$ and $L_{\rm
X}^{2-10\rm\ {keV}}/L_{\rm Edd}$ for the FR Is in our sample is shown
in Figure 3, where the filled circles and empty circles represent
the jet power estimated from the X-ray cavities and the empirical
correlation of Equation 2 respectively. We find that $P_{\rm
jet}/L_{\rm Edd}$ is positively correlated with $L_{\rm X}^{2-10\rm\
{keV}}/L_{\rm Edd}$ for FR Is.
Considering the uncertainties in deriving the jet power and BH mass,
the binned data with standard deviation as errorbar are also plotted
in the figure.

 We calculate the jet power and X-ray luminosity based on
the structure of the ADAFs surrounding Kerr BHs with different
masses, and find that both the jet power and X-ray luminosity are
roughly proportional to the BH mass for $M_{\rm BH}\gtrsim
10^{6}\msun$, provided all other parameters are fixed. We calculate
the $P_{\rm jet}/L_{\rm Edd}$ and $L_{\rm X}^{2-10\rm\ {keV}}/L_{\rm
Edd}$ based on our model varying $\dot{m}$ from $10^{-5}$ to
$10^{-2}$ with $M_{\rm BH}=10^9\msun$ (solid lines in Fig. 3). We
find that the relation between $P_{\rm jet}/L_{\rm Edd}$ and $L_{\rm
X}^{2-10\rm\ {keV}}/L_{\rm Edd}$ calculated with $\alpha=0.3$ and
$\delta=0.3$ can roughly reproduce the observational relation if the
BH spin $j\simeq0.99$ is adopted (see the red solid line in Figure
3). For comparison, the relations for the case of $M_{\rm
BH}=10^8\msun$ is shown as dashed lines, which almost coincide with
the solid lines. Our results suggest that $P_{\rm jet}/L_{\rm
Edd}$-$L_{\rm X}^{2-10\rm\ {keV}}/L_{\rm Edd}$ is insensitive to BH
mass, since that their ratios are roughly scale free from the BH
mass. We further investigate this relation for the case of
$\delta=0.01$ (dotted lines), in which all the other parameters are
unchanged. We find that $j\simeq0.9$ is required in the model
calculations in order to reproduce the observational relation in
this case (blue dotted line in Figure 3). The physical reason is
that the X-ray emission from the ADAF with $\delta=0.01$ is much
weaker than that for $\delta=0.3$ provided all other model
parameters are fixed, which is due to much less dissipated energy
heating electrons directly, while the jet power is not sensitive to
the value of $\delta$. Therefore, our results suggest that the BHs
in FR Is are rapidly spinning with $j\gtrsim0.9$ in order to drive
the observed strong jets.

It should be noted that we assume X-ray luminosities of FR Is to be
dominantly from the ADAF in our calculations. We test this issue by
comparing the relations between the hard X-ray luminosities and
narrow-line luminosities ($L_{\rm [O III]}$) for radio galaxy FR Is
and radio quiet Seyferts (jet is weak or absent), as both the
narrow-line regions in FR Is and Seyferts are believed to be
photo-ionized by the photons emitted from the accretion disks
\citep*[e.g.,][]{wi99}. The results are plotted in Figure 4, in
which FR Is of both part I (filled circles) and part II (open
circles) of our sample are included, and 17 radio quiet,
Compton-thin Seyferts are selected from \citet*[][]{pa06}.
We find that $L_{\rm X}^{2-10\rm\ {keV}}-L_{\rm [O
III]}$ correlation of FR Is is roughly consistent with that of radio
quiet Seyferts.
This means that the X-ray emission of the FR Is and Serferts should
have the same origin. Therefore, the X-ray emission of FR Is should
be mainly from the accretion flows.

\section{Discussion}

We investigate the accretion/jet activities surrounding a spinning
BH, where the accretion flow is described by the ADAF model. The
ADAF may have winds and a power-law radius-dependent accretion rate,
$\dot{m}(r)\propto r^{-p_{\rm w}}$ ($p_{\rm w}>0$) is usually
assumed, although the detailed physics is still unclear
\citep*[][]{bb99}. Therefore, the Bondi accretion rate may be not a
reliable rate of the mass eventually flows into the BH due to the
possible wind, even if it can be estimated from the X-ray
observations. As most of X-ray emission comes from the inner region
of the accretion flow near the BH, the spectrum of ADAF is similar
to that of a ADAF with winds, provided they have the same accretion
rate at their inner edges. As both the X-ray emission and the jets
are from the inner region of the accretion flow, the main conclusion
will not be altered even if the winds are included in the ADAF
model.

The 151 MHz radio luminosity is commonly used to estimate the jet
power for radio galaxies using the relation proposed by
\citet*[][]{wi99}, due to the low-frequency radio emission always
dominated by the extended radio lobes which is free from Doppler
boosting. However, the relation proposed by \citet*[][]{wi99} is
only calibrated for high-power FR IIs, and it is still unclear if it
can be simply extrapolated to low-power FR Is. The X-ray cavities in
galaxies or clusters provide a way to estimate the jet power.
The relation between jet power estimated from the X-ray
cavities, and 327 MHz, 1400 MHz, 5000 MHz, and the radio bolometric
luminosity have been investigated by different groups
\citep*[e.g.,][]{bi04,mh07,bi08,ca10}. \citet*[][]{bi08} proposed
that the radio-source aging may be responsible for some of the
scatter in the relation between jet power and radio luminosity due
to the possible variation of radiative efficiency of the jet, and
the scatter is reduced by $\sim50\%$ after correcting the effect of
radio aging. For example, the relation between jet power and the
radio luminosity becomes tighter when only considering the
radio-filled sources where the jet is still active and particle
injection is occurring \citep*[see][for more details]{bi08}.
However, there is only 5 radio-filled sources used to build the jet
power and radio bolometric luminosity relation in \citet*[][]{bi08},
which is also affected by errors in measuring the radio bolometric
luminosity. In this work, we add 7 more radio sources (FR I or low
power FR II) to investigate the relation between the jet power and
151 MHz radio luminosity, where most of the radio sources have
relatively strong 1400 MHz radio emission which should belong to
radio-filled sources. Compared with the relations in
\citet*[][]{bi08} and \citet*[][]{ca10}, our relation predicts a
little bit lower jet power at given radio luminosity, which is due
to the fact that only the cavities with strong radio emission (most
are FR Is) are considered, while the previous works even included
the ghost cavities, where the radio emission has faded. The
low-frequency radio emission from lobes at 151 MHz does not suffer
from the Doppler boosting effects compared the radio-core emission
used in \citet*[][]{mh07}. Our relation (Equation 2) should be more
accurate for estimate of the jet power for the sources with
relatively strong radio emission (e.g., radio-filled sources or FR
Is in this work). The intrinsic scatter is $\sigma=0.23$ dex for the
$P_{\rm jet}-L_{\rm 151MHz}$ relation derived in our sample, which
is similar to that of 0.3 dex for five radio-filled sources used in
\citet*[][]{bi08}. The scatter of the relation for these
radio-filled sources is much smaller than those of
\citet*[$\sigma\simeq0.47$ dex,][]{mh07} ,
\citet*[$\sigma\simeq0.80$ dex,][]{bi08}, or
\citet*[$\sigma\simeq0.70$ dex,][]{ca10}, in which the radio core
emission is used or ghost-cavities are included. We note that the FR
Is with observed X-ray cavities are still limited, and more
observations are desired for further investigations on this issue.

We employ our ADAF-jet model for Kerr BHs to constrain the BH spins
of FR Is based on the relation between $P_{\rm jet}/L_{\rm Edd}$ and
$L_{\rm X}^{2-10\rm\ {keV}}/L_{\rm Edd}$. For the typical viscosity
parameter $\alpha=0.3$, we find that $j\simeq0.99$ is required to
reproduce the observed relation for the case of $\delta=0.3$, or
$j\simeq0.9$ for the case of $\delta=0.01$. Our results suggest that
the BHs in FR Is should be rapidly spinning with $j\gtrsim0.9$, even
if the uncertainty of the poorly constrained parameter $\delta$ is
considered (see Figure 3). The BH mass used in this work is
mainly derived from the velocity dispersion of the host bulge.
\citet*[][]{la07} argued that the galaxy luminosity may be a better
tracer of the BH mass than the velocity dispersion for the most
massive BHs, and the uncertainties in the high mass end of the
$M_{\rm BH}-\sigma_{*}$ relation can lead to
 underestimation of BH mass up to a factor of several. If this is the
 case, the observational data points will move from the top-right to
 the bottom-left along the 45 degree line in Figure 3, and our main
  conclusion is almost unaltered compared with the model calculations.
   \citet*[][]{da11} found a strong correlation between BH spin and
BH mass, where the spin $j\sim0$ for low mass BHs with $M_{\rm
BH}\sim10^7 \msun$, while it is fast rotating with $j\sim1$ for
supermassive BHs with $M_{\rm BH}\sim10^9 \msun$, based on the
estimation of radiative efficiency for a sample of quasars
\citep*[see][for a similar conclusion]{si07,la09,fa11}. Our
conclusion for the fast rotating BHs in FR Is is similar to their
conclusion, since that the typical BH mass of FR Is is around $10^9
\msun$.

We assume that most of the X-ray
 emission of FR Is is from the ADAFs
in this work. It should be noted that the X-ray emission of some FR
Is with very low Eddington ratios ($L_{\rm X}^{2-10\rm\ {keV}}/\it
L_{\rm Edd}$) may be dominated by the jet emission
\citep*[e.g.,][]{wu07}. \citet*[][]{yc05} also pointed out that the
X-ray emission should be dominated by the emission from the ADAFs
when the Eddington ratio of X-ray luminosity larger than a critical
value and jet contribution will be dominant if the ratio is less
than the critical value. The critical ratio $L_{\rm X,\
crit}^{2-10\rm\ {keV}}/\it L_{\rm Edd}\sim\rm 10^{-7}$ if the typical
BH mass $M_{\rm BH}\simeq10^{9}\msun$ is assumed for FR Is. We find
that most of FR Is in our sample have $L_{\rm X}^{2-10\rm\ {keV}}/\it
L_{\rm Edd}\gtrsim\rm 10^{-7}$ (see Figure 3), and their X-ray
emission should be dominantly from ADAFs considering the criterion
suggested by \citet*[][]{yc05}. This is consistent with our results
based on $L_{\rm X}^{2-10\rm\ {keV}}-L_{\rm [O III]}$ relation for FR
Is and Seyferts. Our main conclusions will not be changed even if
the jets contribute part of the X-ray emission, since the intrinsic
X-ray emission from ADAFs will be less than the observed values
(points in Figure 3 move to left). This will require even higher BH
spins in our model calculations of jet power .

The mass accretion rates of FR Is can be derived from their
X-ray luminosity with our ADAF model calculations for measured BH
masses. We find that the accretion rates are in the range of $\sim
10^{-4}-10^{-2}\dot{M}_{\rm Edd}$ for the FR Is in our sample
  if $\delta=0.3$ is adopted. The BH spins
 of FR Is can be constrained from the jet power and the derived accretion rates (from their
 X-ray luminosities) based on the hybrid jet model \citep*[][]{me01}.
In this work, the jet power is calculated with the mass accretion
rate inferred with the X-ray emission from the ADAF, which should be
better than that estimated from the Bondi accretion rate
\citep*[e.g.,][]{ne07}, since that it may be not the real mass
accretion rate that eventually flowing into BHs due to the possible
outflowing winds or the gas released by the stellar population
inside the
 Bondi radius \citep*[e.g.,][]{pe05,wu07,mc09,mc10}.
  The magnetic field strength constrained from
 the Eddington accretion rate should be upper limits since that the real accretion rate
 should be much less than the Eddington rate in the low power radio sources of their sample,
 and, therefore, the BH spins should be the lower limits \citep*[][]{da09}.
 \citet*[][]{wu08} proposed that the dividing line of FR I/II dichotomy can be well reproduced if
 the BHs are fast rotating with $j\sim0.9-0.99$ and the putative dimensionless accretion
 rate $\dot{m}\sim0.01$ for accretion mode transition is adopted. We calculate the
 accretion rate from the X-ray activities for FR Is in this work, and evaluate the BH spins
 from their jet power and magnetic field strength near the BHs based on the global solution of ADAFs.
 We find that the BHs should be rapidly rotating in FR Is, which further strengthen our
 conclusion of \citet*[][]{wu08}.

The accretion rates of low-power FR Is are very low compared other high-power
FR IIs, which suggest that the accretion disk in FR Is is most probably in an
ADAF state. The ADAF radiates inefficiently and most of gravitational
energy of the accreting matter released is not dissipated locally but advected inward,
and the advected energy may be taken away by the jet or eventually carried into
 the BH. The energy advection plays an important role for jet formation, since that most
 of the gravitational energy heat the protons/electrons and lead to a hot thick disk, which
 allows high poloidal magnetic field strength in the inner region of the flow
 \citep*[e.g.,][]{li99}. \citet*[]{me01} further proposed that the magnetic field
produced by the dynamo process in the ADAF can be amplified due to the frame dragging in
the Kerr metric. We estimate the jet power extracted from the inner region of the ADAF surrounding
a Kerr BH (see Sect. 3), where the field-enhancing shear in the Kerr
metric has been taken into account \citep*[e.g.,][]{me01}. Our global calculations show that the
magnetic field can be amplified around 2 times in the plunging region if the BH spin parameter
$j\simeq0.9-0.99$. Therefore, the jet power was also enhanced roughly 4 times after the
frame dragging effect was considered.  The jet efficiency of $j\simeq0.99$ is around 20\%,
which is 2-3 orders higher than that of low-spin BHs with $j\simeq0$ based on our ADAF-jet
model (see Figure 1).  Our results suggest that a large fraction of the advected energy
can be extracted by the jet if the BH is rapidly spinning (e.g., $j\simeq0.9-0.99$), while
most of the released energy will be advected into the BH if the hole spinning slowly
(e.g., $j\simeq0$). It is interesting to note that the radio
luminosity of radio loud AGNs is always 2-3 orders higher than that
of radio quiet AGNs with similar X-ray or optical luminosities,
which may imply that the BHs have low spins ($j\simeq0$) in radio
quiet AGNs if the jet power is scaled with their radio luminosity.
It is still unclear how to derive the jet power accurately for radio
quiet AGNs (if jets exist in these radio quiet sources), since that
there is no evident extended radio emission as that in radio loud
AGNs, and the detailed calculation for radio quiet sources is beyond
the scope of this paper. Our results roughly support the so-called
spin paradigm that radio loud AGNs have relatively high BH spins,
while radio quiet AGNs have lower BH spins \citep*[e.g.,][and
references therein]{si07,tc10}.

\section*{Acknowledgments}
This work is supported by the NSFC (grants 11143001, 10773020, 10821302, 10833002,
10703009, and 10873005), the National Basic Research Program of China (2009CB824800),
the Science and Technology Commission of Shanghai Municipality (10XD1405000),
the CAS (KJCX2-YW-T03), the Doctoral Program of Higher Education (200804870050),
and the HUST (01-24-012030). This work made use of the HyperLEDA and NED database.

\begin{table}[t]
\footnotesize
  \centerline{\bf Table 1. The data of FR Is.}
  \begin{tabular}{lcccccccc}\hline
Source & $z$ & $\log L_{\rm 151\ MHz}$ & $\log P_{\rm jet}$ & $\log L_{\rm X}^{\rm 2-10\ keV}$& Telescope$^a$ & $\log L_{\rm [O III]}$
& $\log M_{\rm BH}$ & Refs.$^b$\\
         &          & $\rm ergs/s$ & $\rm ergs/s$ & $\rm ergs/s$ &  & $\rm ergs/s$  &  $\msun$     &  \\
\hline
          &          &          &  Part I &         &   &       &       &  \\
3C 84     &   0.018  &   40.86  &   43.88 &   43.40 &  X & 41.60  &  8.64 &  1,1,2,1\\
3C 218    &   0.055  &   42.43  &   44.50 &   42.17 &  C & 40.92  &  8.96 &  1,1,3,1\\
3C 272.1  &   0.004  &   39.15  &   42.70 &   40.34 &  C & 38.20  &  8.80 &  1,1,2,1\\
3C 274    &   0.004  &   40.49  &   43.22 &   40.55 &  C & 38.99  &  9.48 &  1,1,2,1\\
3C 317    &   0.034  &   41.46  &   44.05 &   41.89 &  C & 40.35  &  8.58 &  1,1,2,1\\
3C 338    &   0.030  &   41.26  &   43.64 &   40.56 &  C & 39.57  &  8.92 &  1,1,2,1\\
3C 388    &   0.092  &   41.93  &   44.15 &   41.69 &  C & 40.70  &  9.18 &  1,1,2,1\\
3C 405    &   0.056  &   44.04  &   45.23 &   44.22 &  C & 42.48  &  9.40 &  1,1,4,1\\
NGC 507   &   0.016  &   39.37  &   43.29 &   39.90 &  A &  ...   &  8.90 &  5,1,1\\
NGC 1316  &   0.006  &   38.11  &   42.04 &   39.46 &  C & 38.50  &  8.34 &  5,6,7,8$^c$\\
NGC 4696  &   0.010  &   39.97  &   42.71 &   40.26 &  C & 39.10  &  8.60 &  1,1,3,1\\
IC 4374   &   0.022  &   39.67  &   43.12 &   41.37 &  A & ...    &  8.57 &  1,1,1\\
\hline
          &          &          & Part II &         &    &    &       & \\
3C 15     &   0.073  &   41.66  &   44.06 &   42.40 &  C & 40.60  &  8.70 &  6,2,9 \\
3C 31     &   0.017  &   40.31  &   43.36 &   40.67 &  C & 39.46  &  8.70 &  10,2,8\\
3C 66     &   0.021  &   40.82  &   43.62 &   41.11 &  B & 40.05  &  8.84 &  10,2,11 \\
3C 78     &   0.029  &   40.88  &   43.66 &   42.16 &  C & 39.41  &  8.60 &  6,2,12 \\
3C 83.1   &   0.025  &   40.84  &   43.64 &   41.11 &  C & 39.50  &  9.01 &  10,2,9 \\
3C 189    &   0.043  &   40.85  &   43.64 &   41.79 &  C &  ...   &  8.93 &  11,11 \\
3C 264    &   0.021  &   40.55  &   43.49 &   41.88 &  C & 39.20  &  8.61 &  10,2,8 \\
3C 270    &   0.007  &   40.03  &   43.22 &   41.19 &  C & 38.96  &  8.88 &  13,2,8 \\
3C 278    &   0.015  &   40.41  &   43.41 &   39.70 &  C & 39.40  &  8.62 &  6,3,8 \\
3C 288    &   0.246  &   42.79  &   44.65 &   42.48 &  C & 40.65  &  8.72 &  14,2,8 \\
3C 296    &   0.024  &   40.33  &   43.37 &   41.49 &  C & 39.78  &  8.80 &  10,2,8 \\
3C 305    &   0.042  &   41.08  &   43.76 &   42.04 &  C & 41.03  &  8.10 &  14,2,8 \\
3C 315    &   0.108  &   41.98  &   44.23 &   41.71 &  C & 40.87  &  8.71 &  14,2,15 \\
3C 317    &   0.034  &   41.45  &   43.95 &   41.48 &  C & 40.35  &  8.26 &  10,2,8  \\
3C 346    &   0.162  &   42.16  &   44.32 &   43.28 &  C & 41.24  &  8.89 &  11,2,11 \\
3C 438    &   0.290  &   43.34  &   44.94 &   42.51 &  C & 41.46  &  8.80 &  11,2,11 \\
3C 449    &   0.017  &   40.15  &   43.28 &   40.46 &  X & 39.19  &  8.54 &  11,2,8 \\
3C 465    &   0.030  &   41.15  &   43.80 &   41.11 &  X & 39.81  &  9.13 &  10,2,8 \\
NGC 6109  &   0.030  &   40.17  &   43.29 &   40.36 &  C & 39.71  &  8.56 &  10,16,8 \\
NGC 6251  &   0.025  &   40.66  &   43.54 &   42.78 &  C & 41.77  &  8.97 &  10,4,8 \\
NGC 315   &   0.017  &   39.74  &   43.07 &   41.71 &  C & 40.87  &  8.71 &  11,4,8 \\
Cen A     &   0.002  &   39.39  &   42.88 &   41.70 &  C & 38.82  &  8.54 &  10,4,12 \\
IC 4296   &   0.013  &   39.93  &   43.17 &   41.17 &  C & 39.50  &  9.01 &  6,3,8 \\

\hline
\end{tabular}

\begin{minipage}{170mm}
$^a$ Telescope with which the object was observed: X=$XMM$-$Newton$; C=$Chandra$; A=$ASCA$; B=$BeppoSAX$. \\
$^b$ The references for the jet power, X-ray luminosity, [O III] luminosity and BH mass respectively (Part I),
while jet power for sample of part II is calculated in this work.\\
$^c$ The BH mass is calculated from the $M_{\rm BH}-\sigma$ relation in this work.\\
References: (1) \citet*[][]{mh07}; (2) \citet*[][]{bu10}; (3) \citet*[][]{wi04};
(4) \citet*[][]{ba99}; (5) \citet*[][]{ca10}; (6) \citet*[][]{ri05};
(7) \citet*[][]{si96}; (8) this work; (9) \citet*[][]{ma04};
(10) \citet*[][]{ev06}; (11) \citet*[][]{do04}; (12) \citet*[][]{ba08};
(13) \citet*[][]{ze05}; (14) \citet*[][]{ha09}; (15) \citet*[][]{cr04};
(16) \citet*[][]{wo07}.

\end{minipage}

\end{table}


\clearpage

\begin{figure}
\epsscale{1.0} \plotone{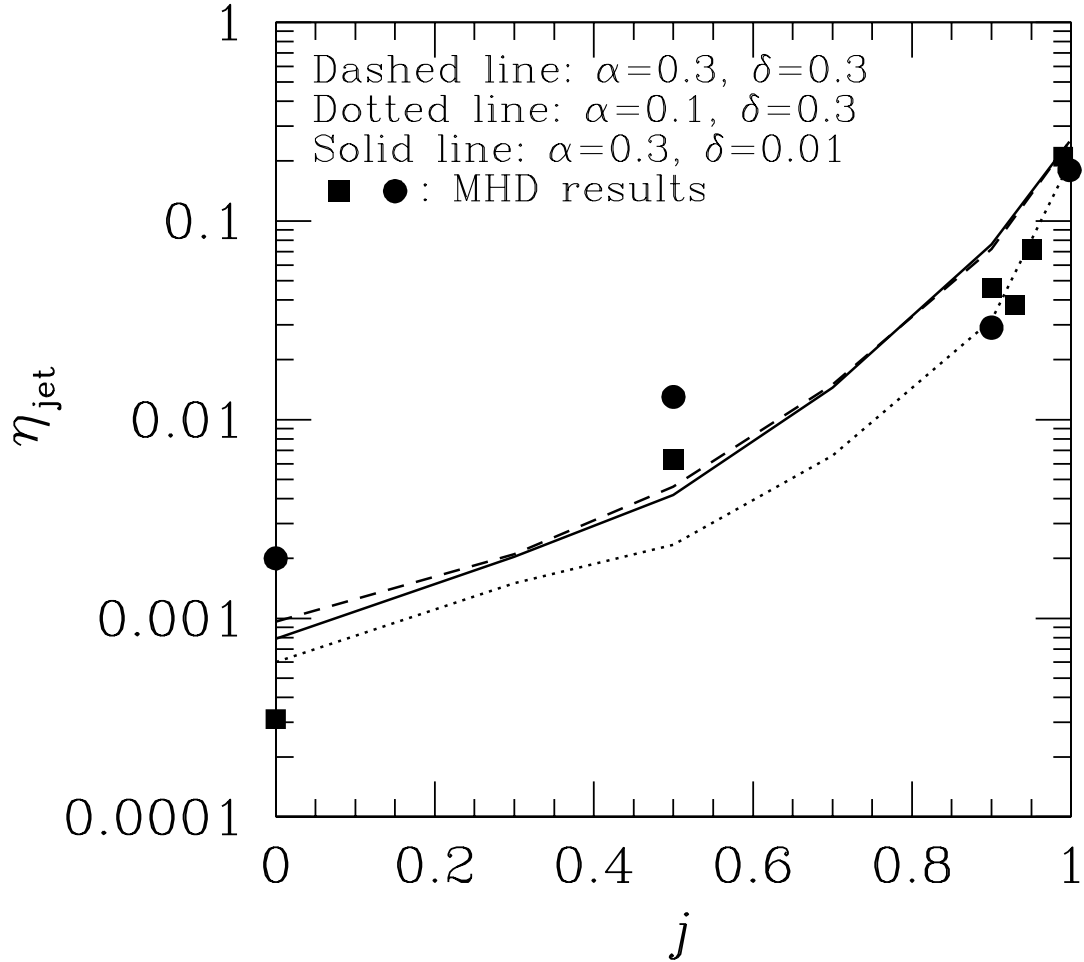} \caption{The jet efficiency,
$\eta_{\rm jet}=P_{\rm jet}/\dot{M}c^{2}$ , predicted by our ADAF
model calculations. The dashed line and dotted line show the results
for $\alpha=0.3$ and $\alpha=0.1$ with $\delta=0.3$ respectively,
while the solid line represents the case of $\alpha=0.3$ with
$\delta=0.01$. Estimates of $\eta_{\rm jet}$ from the numerical
simulations of \citet*[][]{de05} and \citet*[][]{hk06} are shown by
the filled circles and squares respectively.
  \label{fig1}}
\end{figure}

\begin{figure}
\epsscale{1.0} \plotone{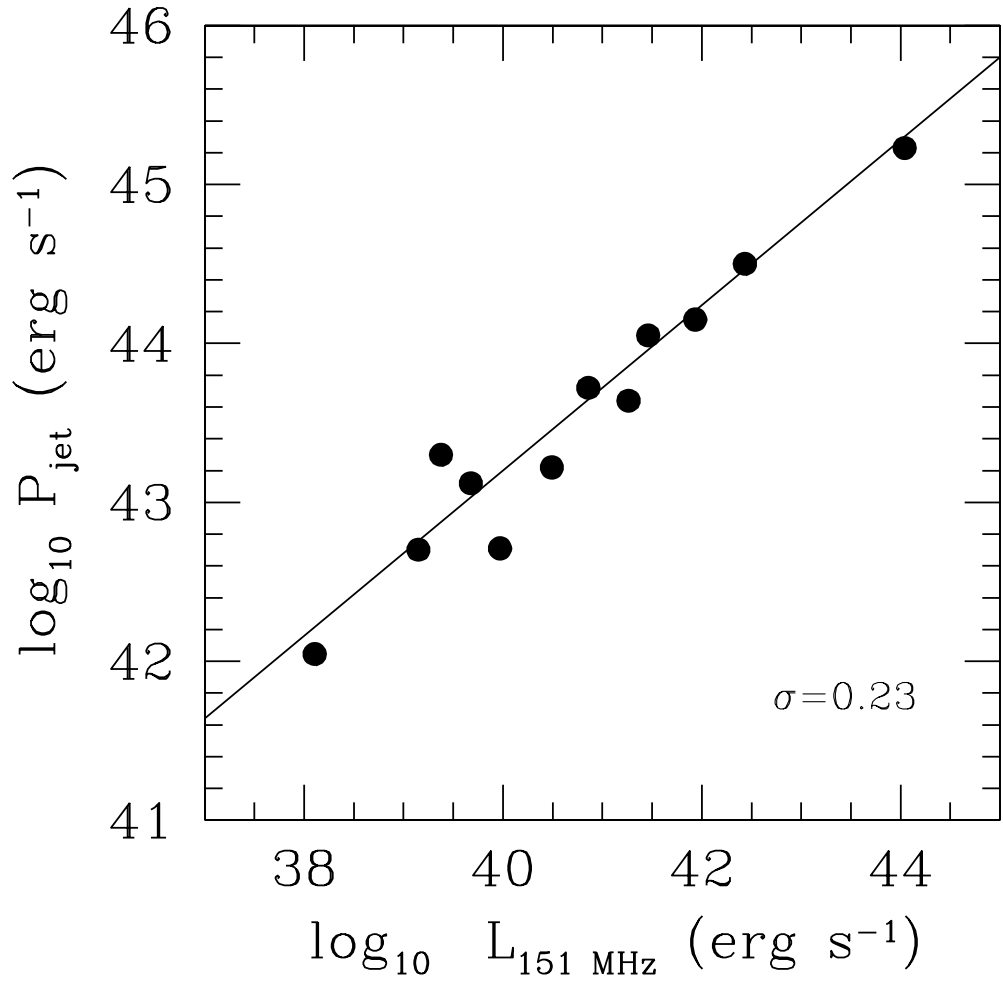} \caption{The relation between jet power and
 151 MHz luminosities for the sources with jet power estimated from their X-ray cavities.
 The solid line is the best fit of the data.  \label{fig2}}
\end{figure}

\begin{figure}
\epsscale{1.0} \plotone{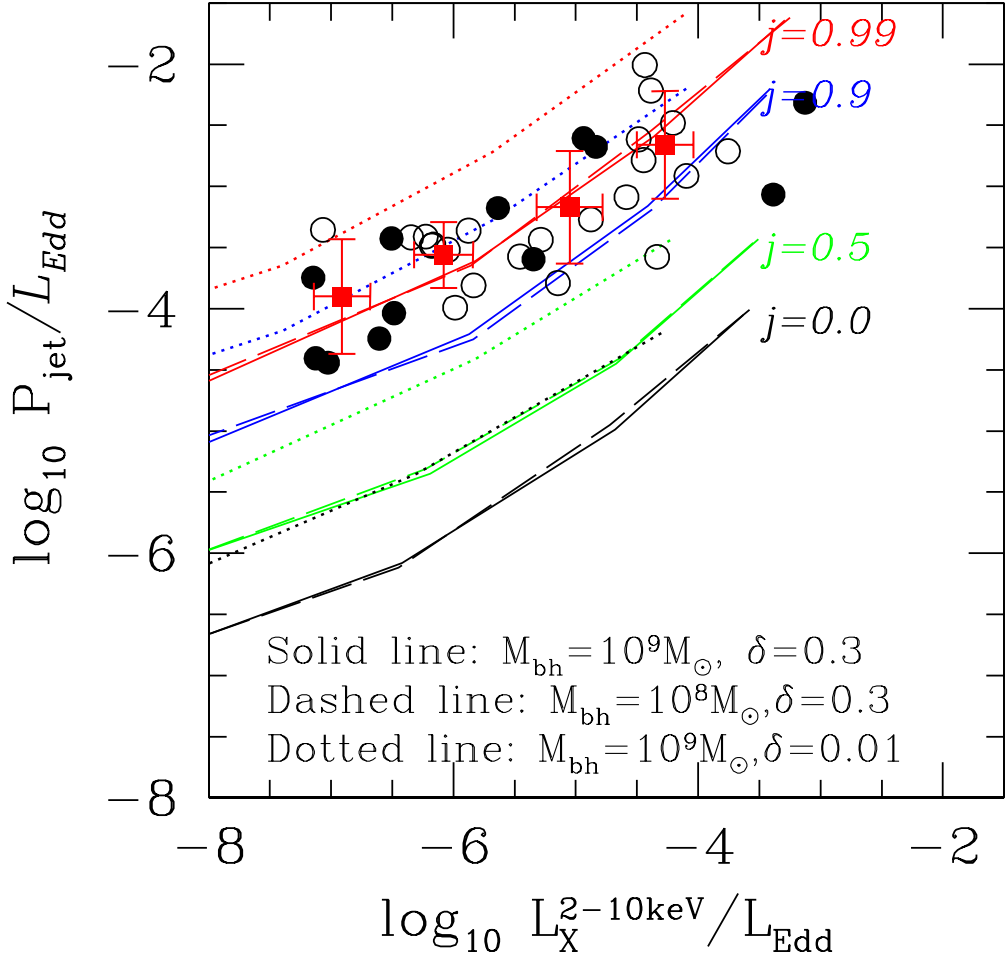} \caption{$P_{\rm jet}/L_{\rm
Edd}$ versus $L_{\rm X}^{2-10\ {\rm keV}}/L_{\rm Edd}$. The filled
circles represent FR Is with jet power estimated from X-ray
cavities, and the jet power of FR Is estimated with the empirical
correlation of Eq. 2 are marked with open circles. The result for
the sources binned in the $L_{\rm X}^{2-10\ {\rm keV}}/L_{\rm Edd}$
interval of [-7.5,-6.5], [-6.5,-5.5], [-5.5,-4.5], [-4.5,-3.5], is
shown as red squares with standard deviation. The solid lines and
the dashed lines represent the model predictions with $M_{\rm BH}=10^9\msun$
and $10^8\msun$ for the case of $\delta=0.3$ respectively ,where the red,
blue, green and black colors represent $j=$ 0.99, 0.9, 0.5 and 0, respectively
 (from top to bottom). The dotted
lines are for the case of $\delta=0.01$ with $M_{\rm BH}=10^9\msun$.
 \label{fig3}}
\end{figure}

\begin{figure}
\epsscale{1.0} \plotone{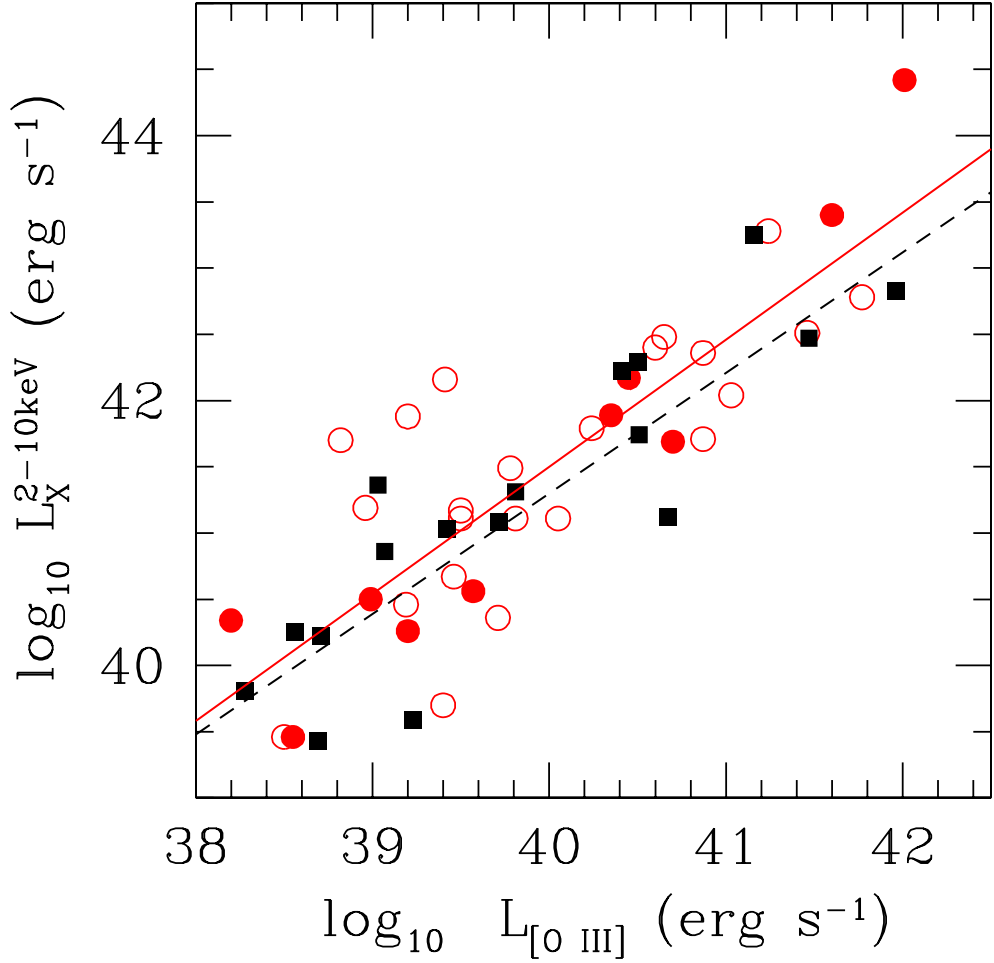} \caption{The relation between
$L_{\rm X}^{2-10\ {\rm keV}}$ and $L_{\rm [O\ III]}$. The filled
circles and open circles represent the jet power of FR Is estimated
with the X-ray cavities and the empirical relation in this work,
respectively. The black filled squares represent radio quiet
Seyferts, which are taken from \citet*[][]{pa06}. The solid line and
dashed line are the best fits of FR Is and Seyferts, respectively.
   \label{fig3}}
\end{figure}

\end{document}